\documentclass[10pt,letterpaper]{article}
\usepackage[top=0.85in,left=2.75in,footskip=0.75in]{geometry}

\usepackage{amsmath,amssymb}

\usepackage{changepage}

\usepackage{textcomp,marvosym}

\usepackage{cite}

\usepackage{nameref,hyperref}

\usepackage[nopatch=eqnum]{microtype}
\DisableLigatures[f]{encoding = *, family = * }

\usepackage[table]{xcolor}

\usepackage{array}

\newcolumntype{+}{!{\vrule width 2pt}}

\newlength\savedwidth

\newcommand\thickhline{\noalign{\global\savedwidth\arrayrulewidth\global\arrayrulewidth 2pt}%
\hline
\noalign{\global\arrayrulewidth\savedwidth}}

\raggedright
\usepackage[aboveskip=1pt,labelfont=bf,labelsep=period,justification=raggedright,singlelinecheck=off]{caption}

\makeatletter
\renewcommand{\@biblabel}[1]{\quad#1.}
\makeatother

\usepackage{lastpage,fancyhdr,graphicx}
\usepackage{epstopdf}
\fancyheadoffset[L]{2.25in}
\fancyfootoffset[L]{2.25in}
\begin{document}
\vspace*{0.2in}

\begin{flushleft}
{\Large
\textbf\newline{A local eigenvector centrality} 
}
\newline
\\
Ruaridh A. Clark\textsuperscript{1*},
Francesca Arrigo\textsuperscript{2},
Agathe Bouis\textsuperscript{1},
Malcolm Macdonald\textsuperscript{1}
\\
\bigskip
\textbf{1} Centre for Signal and Image Processing, University of Strathclyde, UK
\\
\textbf{2} Department of Mathematics and Statistics, University of Strathclyde, UK
\\
\bigskip

* ruaridh.clark@strath.ac.uk

\end{flushleft}
\section*{Abstract}
Eigenvector centrality is an established measure of global connectivity, from which the importance and influence of nodes can be inferred. We introduce \textit{local eigenvector centrality} (LEC), a scale-adaptive extension that integrates local and global connectivity into a single centrality measure. LEC identifies a topologically relevant mesoscopic scale for community-level centrality assessment, informed by prominent eigengaps, and combines information from the associated eigenvectors, via the Euclidean norm, to yield a centrality that reflects community-driven node influence. Networks with known community structures are employed to demonstrate that LEC closely reproduces eigenvector centrality for each community in isolation. Discrepancies between these measures are shown to be attributable to the global connectivity of affected nodes. LEC therefore provides a principled integration of community-level and global notions of node importance. Comparisons with PageRank and Katz centrality demonstrate that damping-based approaches do not recover community-scale centrality as effectively as LEC.

\section*{Author summary}
Eigenvector centrality is a long established measure of node importance for systems that are driven by global connectivity. For systems where local, community-level, connections and process are dominant we introduce a \textit{local eigenvector centrality} (LEC) to identify the dominant scale for measuring centrality and reflect the importance of nodes in driving both local and global processes. LEC is shown to be a distinctive and effective measure for accurate and consistent mapping of community-level node centrality.


\section*{Introduction}
Eigenvector centrality is a long-established network measure that quantifies node importance in terms of global reach and influence \cite{Bonacich1972,Bonacich2007}. It is defined by the principal eigenvector of a network's adjacency matrix $A$, corresponding to the largest eigenvalue, such that a node's centrality is intertwined with that of its neighbours. Consequently, eigenvector centrality provides a recursive measure of importance that is particularly appropriate when influence is dominated by highly connected hubs or when dense global connectivity diminishes the relevance of local structure. However, many real-world networks -- across social, biological and technological systems -- exhibit pronounced mesoscopic organisation in which local structures and communities play a central role in shaping global dynamics. Examples range from disease transmission through contact networks \cite{Clark2021,Sah2017,SalathM2010}, collective coordination in starling flocks \cite{Clark2019}, power distribution across Europe \cite{SchaubMichaelT2012}, and consensus formation in satellite constellations \cite{bouis2026insights}.

We introduce \textit{local eigenvector centrality} (LEC), a measure that determines node importance at the most topologically relevant scale through the identification of prominent eigengaps. In doing so, LEC adapts to the network's mesoscopic organisation to better characterise node centrality at a local scale. LEC extends Bonacich’s eigenvector centrality \cite{Bonacich1972} by integrating information from a subset of eigenvectors associated with a prominent eigengap. In this respect, LEC is conceptually related to eigengap-aware spectral clustering methods that embed a network in a Euclidean space defined by the graph’s eigenvectors, with algorithms such as k-means being applied to detect node clusters \cite{VonLuxburg2007,Shen2010}. Whereas spectral clustering aims to partition a graph into discrete communities, LEC quantifies node importance by accounting for both global and community connectivity without requiring explicit community assignments. 

Spectral embeddings and eigengap analysis have also been employed across a range of studies where the choice of network matrix representation has been shown to be influential. In \cite{Shen2010}, the normalised Laplacian was identified as the most effective for determining community presence from its eigengaps, following comparisons with adjacency, Laplacian, modularity and correlation matrices. Similarly in Ref. \cite{Kojaku2024} the most effective machine learned embeddings for community detection were found to closely resemble the spectral embeddings generated from the normalised Laplacian matrix. The adjacency matrix has also been applied to perform clustering in reference to the largest eigengap, where node correlations from an extended set of eigenvectors determined partitions \cite{Luciska2018}. Herein, the adjacency matrix is used to identify centrality in terms of node reach and connectivity which is distinct from the notion of leadership and dynamical influence for which the Laplacian is well-suited \cite{Clark2019,Punzo2016}.

There exists a plethora of centrality measures, including community-based, local, semi-local and neighbourhood measures that appear conceptually related to LEC \cite{shvydun2025zoo}. However, such measures often quantify distinct notions of importance and therefore serve different analytical purposes to LEC. For example, community-aware centrality measures -- requiring \textit{a priori} community knowledge or community detection -- frequently seek to identify nodes that bridge between communities or facilitate inter-community communication \cite{rajeh2023comparative}. In contrast, LEC seeks to characterise centrality at local mesoscopic scales of network organisation, such that known communities can be used to validate the resulting centrality assessments.

LEC is conceptually related to random-walk and diffusion-based centrality measures. The most widely known example is PageRank \cite{Page1999}, which can be conceptualised as assessing the popularity of nodes as destinations for random walkers traversing a graph. At each node, the probability of traversing an outgoing edge is determined by the relative edge weights of the available connections. Eigenvector centrality admits a similar interpretation, although node popularity propagates according to absolute edge weights rather than transition probabilities. Both measures can be affected by localisation, whereby centrality becomes concentrated around densely connected hubs or isolated subgraphs, analogous to random walkers becoming trapped within a restricted region of the network and repeatedly visiting the same nodes \cite{Pastor-Satorras2016}. PageRank -- with its random teleportation (damping) factor -- and the non-backtracking centralities \cite{Arrigo2020,Pastor-Satorras2016} -- that prevents return along traversed edges -- have been developed in part to mitigate localisation in the calculation of centrality. LEC is not developed to avoid the issue of localisation; nevertheless, a mitigation strategy is presented herein. 

Katz centrality is another random-walk-based measure that employs a damping factor to both mitigate localisation and tune the relative importance of short and long paths through a graph. Consequently, Katz centrality accounts for both local connectivity and broader network structure, a perspective that is conceptually related to LEC in seeking to balance community-level importance with global reach. However, damping-based spectral centrality measures have been shown to be equivalent to the dominant eigenvectors of suitably perturbed graphs \cite{vigna2016spectral}. Hence, the resulting centrality remains driven by dominant spectral modes. By contrast, LEC explicitly exploits information contained within a broader set of eigenvectors to investigate mesoscopic structural features that may not be apparent from dominant eigenmodes alone.

Principal Component Centrality (PCC) is a centrality measure that similarly exploits the eigenspace embedding of a graph and incorporates information from multiple eigenvectors \cite{Ilyas2010}. PCC differs from LEC in two important respects. First, PCC is defined as the Euclidean norm of the vector formed by weighting each of the selected eigenvector components by their corresponding eigenvalue. Thereby, placing greater emphasis on the dominant spectral modes and favouring global assessments of centrality. Second, the set of eigenvectors retained is typically determined using a phase-angle convergence heuristic analogous to variance-explained criteria in principal component analysis. Eigenvectors are successively incorporated until the phase angle between the resulting PCC vector and the eigenvector centrality vector stabilises, indicating that additional eigenvectors produce little further change in the PCC solution. In contrast, LEC selects eigenvectors according to prominent eigengaps, allowing the analysis scale to be tuned to various mesoscopic structural features of the network. Consequently, LEC is designed to identify node importance within the most topologically relevant spectral subspace, with all retained eigenvectors contributing equally to the centrality measure. This distinction enables LEC to characterise community-level and multi-scale notions while incorporating, but not being dominated by, the leading eigenmodes of the graph.

\section{Methodology}
In this section the local eigenvector centrality (LEC) measure is defined alongside a series of associated hypotheses.

\subsection{Local eigenvector centrality}
An adjacency matrix, $A$, is a square $n \times n$ matrix representing a graph of $n$ vertices, where $a_{ij} > 0$, with $a_{ij}$ the $(i,j)$-th entry of the graph’s adjacency matrix if there exists a directed edge from vertex $i$ to $j$, and $0$ otherwise.

LEC is defined in reference to a prominent real eigengap among the eigenvalues $\lambda_1, \lambda_2, \dots, \lambda_n \in \mathbb{C}$ of $A$, where we assume the ordering  
\begin{align}\label{2.1}
\mathrm{Re}(\lambda_1) \geq \mathrm{Re}(\lambda_2) \geq \dots \geq \mathrm{Re}(\lambda_n).
\end{align}

The above ordering provides a real-valued spectral hierarchy from which eigengaps can identify mesoscopic network structure. Complex eigenvalues can occur, and appear in conjugate pairs since $A$ is real. In reference to the geometric interpretation of complex eigenvalues given by Block Diagonalization Theorem \cite{MargalitRabinoff2019}, 
the imaginary component captures rotation and the real component the scaling of $A$.
In contrast to a real-valued spectral heirarchy, ordering by the modulus can artificially promote eigenvalues with large imaginary components that do not correspond to prominent network structures, see \nameref{S1_Appendix}.

The local eigenvector centrality is based on a prominent, $i$-th eigengap, where  
\begin{align}\label{2.2}
g_i = \mathrm{Re}(\lambda_{i}) - \mathrm{Re}(\lambda_{i+1}).
\end{align}
For real eigenpairs, the sign of an eigenvalue determines whether the associated eigenvector direction is preserved or reversed, while its magnitude determines the scaling. Accordingly, we restrict the eigengap search to eigenvalues with non-negative real parts, denoted as  
\begin{align}\label{2.3}
k = \operatorname*{arg\,max}_{\mathrm{Re}(\lambda_{i+1}) \ge 0} g_i.
\end{align}


To define the LEC measure, we introduce a variant of the matrix of eigenvectors to accommodate complex eigenpairs. Recall that each distinct eigenvalue $\lambda$ of $A$ is associated with at least one eigenvector $\textbf{v} \neq 0$, satisfying $A \textbf{v} = \lambda \textbf{v}$. Whenever $\lambda$ is a complex eigenvalue of $A$, its complex conjugate $\lambda^*$ is also an eigenvalue, i.e.,  
\begin{align}\label{2.4}
A \textbf{v} = \lambda \textbf{v} \quad \Longleftrightarrow \quad A \textbf{v}^* = \lambda^* \textbf{v}^*.
\end{align}
Moreover, because of the above ordering, conjugate pairs of eigenvalues correspond to consecutive indices in our list.
Given that a complex eigenvector $\textbf{v}=\textbf{u}+i\textbf{w}$ is composed of two real vectors $\textbf{u}$ and $\textbf{w}$, that form an invariant two-dimensional subspace $S=span\{\textbf{u},\textbf{w}\}$
\cite{MargalitRabinoff2019} and \cite[Prop.~6.4.4]{tudelft_complex_eigenvalues}, we replace complex eigenvectors when building the matrix  
\begin{align}\label{2.6}
V = [\textbf{v}_1 \;|\; \textbf{v}_2 \;|\; \dots \;|\; \textbf{v}_k] \in \mathbb{R}^{n \times k}
\end{align}
according to the following rules:
\begin{itemize}
    \item If $\lambda_i \in \mathbb{R} \setminus \{0\}$, then $\textbf{v}_i \in \mathbb{R}^n$ satisfies $A \textbf{v}_i = \lambda_i \textbf{v}_i$ and $\|\textbf{v}_i\|_2 = 1$.
    \item If $\lambda_i\in\mathbb{C}\setminus\mathbb{R}$, then $\lambda_{i+1} = \lambda_i^*$ and  
    \begin{align}\label{2.7}
    \textbf{v}_i = \frac{\textbf{u}}{\|\textbf{u}\|_2} , 
    \quad 
    \textbf{v}_{i+1} = \frac{\textbf{w}}{\|\textbf{w}\|_2},
    \end{align}
    where $\textbf{v}_i\in\mathbb{C}^n$ is the eigenvector associated with $\lambda_i$ with $\|\textbf{v}_i\|_2 = 1$.
    \item If $\lambda_i = 0$, then $\textbf{v}_i = 0$.
\end{itemize}
These rules ensure that for each non-zero \textit{real} eigenvalue, the corresponding eigenvector can be incorporated into LEC directly. For each conjugate pair, we replace the associated complex eigenvector by its normalised real and imaginary components, which form a real basis for the corresponding invariant subspace. Finally, eigenvalues equal to zero contribute a zero vector. Hence, although $V$  is derived from the eigenspace of $A$ , it is not itself a matrix of eigenvectors. This distinction is further justified in \nameref{S1_Appendix}. 

With this notation, the vector of local eigenvector centralities $c$ is defined entrywise as  
\begin{align}\label{2.8}
c(i) = \sqrt{\sum_{j=1}^k v_j(i)^2} 
= \|[v_1(i), \dots, v_k(i)]\|_2 
= \| e_i^\top V \|_2,
\end{align}
where $e_i$ is the $i$-th standard basis vector of $\mathbb{R}^n$, with $e_i(i) = 1$ and $0$ elsewhere.

The behaviour of LEC in representative edge cases, including directed paths and directed cycles, is examined in \nameref{S1_Appendix}.

\textbf{Remark:} The above construction implicitly assumes that all the eigenvalues of $A$ up to the $k$-th are semi-simple, i.e., that their algebraic and geometric multiplicities coincide. This is not always the case, as eigenvalues may be repeated $s$ times in the spectrum but have $<s$ associated linearly independent eigenvectors. In such scenarios (when an eigenvalue is defective), we only consider eigenvectors and disregard generalised eigenvectors arising from Jordan chains; see, e.g., \cite{meyer2023matrix,dym2013linear}. 

\subsubsection{Eigengaps and community structure}
The role of eigenvalues and their eigengaps in identifying prominent community structures can be understood through the following analogy to a physical system. Consider a spring system composed of two densely connected communities with only a few connections bridging between them. Driving this structure apart, in two separate directions according to the second eigenvector, will achieve a relatively large distance between the two communities. The real part of the second eigenvalue reflects this physical separation with a relatively large value. To drive the same structure in three distinct directions, corresponding to the third eigenvalue, would result in less separation between groups with at least one of the two densely connected communities being split into two sets of nodes. Therefore, the third eigenvalue would have a relatively small real component. This produces a large difference between the magnitude of the real second and third eigenvalue components and therefore a large eigengap that indicates the presence of community structure.

\subsection{Centrality benchmarks}
The most relevant centrality measures for benchmarking the performance of LEC are described below. Additional comparisons with selected community-aware centrality measures \cite{rajeh2023comparative}, Neighbourhood and local centrality measures \cite{shvydun2025zoo} are reported in the \nameref{sec:results} section. These methods are not discussed in detail, as they serve as supplementary benchmarks rather than primary comparators.

\subsubsection{Known-Community Eigenvector Centrality}\label{sec:kcec}
Known communities are those defined with \textit{a priori} knowledge of a graph and its community structure. To provide a community-centric benchmark for comparison with centrality metrics, we define \textit{Known-Community Eigenvector Centrality (KCEC)} as eigenvector centrality computed independently for each known-community's subnetwork. We extract the subnetwork for each community and compute the first, principal, eigenvector of the corresponding subnetwork adjacency matrix. KCEC is presented as a single vector composed of each community's independent centrality assessment.

\subsubsection{Katz Centrality} 
Katz centrality \cite{Katz1953} measures node importance by considering both direct and indirect connections, with contributions from longer walks attenuated by a factor $\alpha$. 
The resulting scores accumulate contributions from all walks terminating at a node, with shorter walks contributing more strongly than longer walks. Unlike eigenvector centrality, Katz centrality assigns positive scores to all nodes and remains well defined even for disconnected components. 

The centrality vector is 
$\mathbf{x} = \alpha A^{\mathsf{T}}\mathbf{x}+\mathbf{1}$,
where $A$ is the adjacency matrix, $\mathbf{1}$ is a vector of ones, $\alpha$ is the attenuation factor satisfying $0<\alpha < 1/|\lambda_{\max}|$, where $\lambda_{\max}$ is the largest eigenvalue of $A$. Rearranging yields 
\begin{align} 
\mathbf{x} = (I-\alpha A^{\mathsf{T}})^{-1}\mathbf{1}
\end{align}
where $I$ is the identity matrix.

\subsubsection{PageRank}
PageRank (PR) \cite{Page1999} incorporates a damping factor $\alpha \in (0,1)$, which models the probability that a random walker follows an edge or teleports to another node as
\begin{align}
\mathbf{x} = \alpha W \mathbf{x} + \frac{(1-\alpha)}{n}\mathbf{1},
\end{align}

\noindent where $\mathbf{x}$, is a vector of PageRank scores,  $\mathbf{1}$ is a vector of ones and $W = \{w_{i,j}\}$ is the column-stochastic transition matrix defined as $w_{i,j} = {a_{j,i}}/{\sum_k a_{j,k}}$.

\subsubsection{Principal Component Centrality} 
Principal Component Centrality (PCC) \cite{Ilyas2010} assigns each node a score equal to the Euclidean norm of its coordinates in a $P$-dimensional eigenspace formed by the leading eigenvectors of the adjacency matrix. Let $\lambda_k$ and $\mathbf{x}_k$ denote the eigenvalue and corresponding eigenvector of the adjacency matrix $A$, ordered by decreasing eigenvalue magnitude. The PCC of node $i$ is 
\begin{align}
\mathrm{PCC}_i = \sqrt{ \sum_{k=1}^{P} (\lambda_k x_{ik})^2 }, 
\end{align}
where $x_{ik}$ is the $i$th element of eigenvector $\mathbf{x}_k$. When $P=1$, PCC reduces to a scaled form of eigenvector centrality. $P$ is selected using a phase-angle stability criterion that identifies the point at which the phase angle between the PCC and EC vectors stabilises as additional eigenvectors are incorporated.

\subsection{Hypotheses} \label{sec:hypotheses}

LEC is designed to integrate information from multiple prominent eigenvectors to capture both global and local community structure in networks. To evaluate the validity of this design, we define the following hypotheses:

\begin{itemize}
    \item \textbf{H1: LEC is community-aware.}\\
In networks with known communities -- that are not composed of prominent sub-communities -- local eigenvector centrality will closely approximate eigenvector centrality computed for each community in isolation. 
    \item \textbf{H2: LEC integrates global and local centrality.}\\
In networks with known-communities -- that are not composed of prominent sub-communities -- nodes with higher LEC than KCEC will exhibit greater global connectivity. 
    \item \textbf{H3: Damping-based locality is not equivalent to community centrality.}\\
 Katz and PageRank use damping factors to attenuate the influence of long walks, thereby increasing the prominence of local neighbourhood structure. However, locality induced by walk-length attenuation is not equivalent to locality induced by mesoscopic community structure. Consequently, node rankings produced by LEC will differ systematically from those produced by Katz and PageRank.
\end{itemize}

\subsubsection{Hypothesis testing} \label{sec:hypo_testing}
Centrality measures are normalised by their sum to remove any differences in scale between measures with Pearson’s correlation used to assess similarity between centrality measures. 

To identify whether LEC reflects global centrality, Spearman rank correlation is used for the difference between LEC and global centrality measures with KCEC. The correlation therefore seeks to capture the agreement in the ordering of node-wise deviations from KCEC, where nodes with higher LEC than KCEC would also have higher global centrality deviation from KCEC.

To enable direct comparison between eigenvector-based centrality measures and PageRank, the adjacency matrix is transformed into a row-stochastic matrix by normalising each row to have a unit $\ell_1$ norm (i.e. row sum to one), an assessment referred to as Random Walk Centrality or Seeley's Index \cite{seeley1949net}. This transformation ensures that the matrix represents probabilistic flow on a directed graph, with the left eigenvectors defining the corresponding eigenvector-based centrality measures to ensure alignment with PageRank. To assess the sensitivity of PageRank to the damping factor $\alpha$, we evaluated \(\alpha\in[0.01,1]\).

To assess the sensitivity of Katz centrality to the damping factor, we evaluated $\alpha=\frac{c}{\lambda_{\max}}$, where \(c\in[0.01,0.99]\). Parameterising \(\alpha\) relative to the spectral radius ensures that the convergence condition \(\alpha<1/\lambda_{\max}\) is satisfied for all networks while allowing comparison across a range of short- to long-walk regimes. Small values of \(c\) emphasise direct neighbourhood structure by heavily down-weighting longer walks, whereas values approaching unity assign greater weight to long walks and produce rankings increasingly similar to eigenvector centrality \cite{benzi2015limiting}.

\subsubsection{Synthetic scenarios}
To test these hypotheses, synthetic networks are generated as undirected, unweighted graphs with known-community structure and tunable parameters. Each network is comprised of six topologically identical communities, with twenty nodes per community. The base community topology is constructed from a randomly ordered, undirected, spanning tree to ensure all community nodes are connected. Additional intra-community edges are added with probability of 0.15 to increase community connectivity without adding self-loops.

Inter-community connections are added to each of the six communities to create variation in connectivity patterns. The number of inter-community connections defined the number of nodes randomly assigned a pairwise connection with another community node. Consequently, increasing the number of inter-community connections increases the variation in connectivity patterns between the six communities.

A random seed is specified to enable reproducibility of the network topology and node selection. The resulting networks therefore provide a controlled framework in which inter-community connectivity is varied for a set of synthetic known-community networks.

\section{Results}\label{sec:results}

LEC is proposed to more accurately capture the distribution of node centrality in networks with prominent community structure. LEC is evaluated on synthetic and real-world data where known community structures supports the assessment of hypotheses H1--H3.

\subsection{Synthetic networks}

\begin{figure}[t!]
    \centering
    \includegraphics[width=1\textwidth]{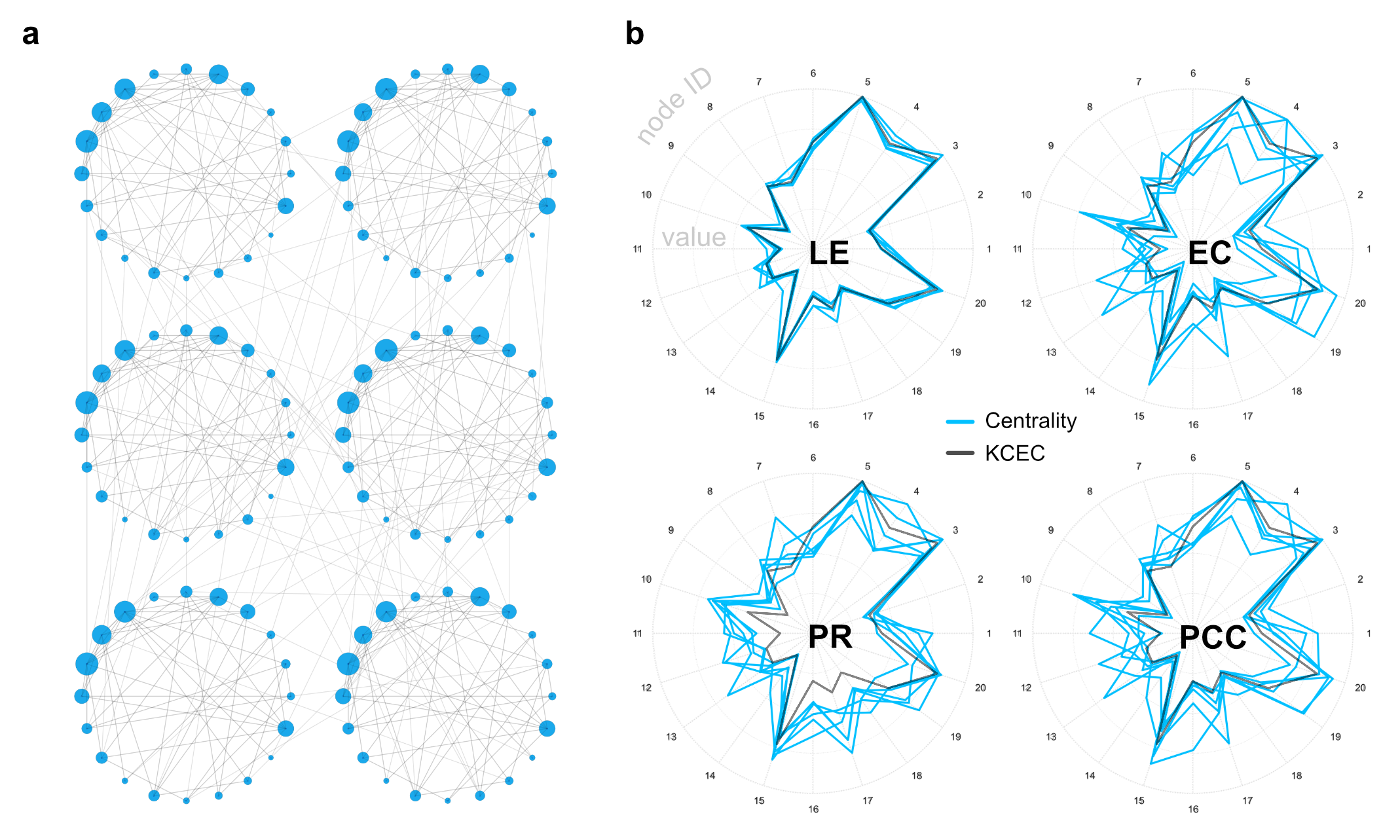}
    \caption{{\bf Centrality analysis of synthetic graph comprising six duplicated communities with randomly allocated edges connecting communities.} (a) LEC values represented by node marker area. (b) Polar plots of LEC, EC, PageRank (PR) and PCC allocations for the 20 nodes in each community. KCEC values are overlaid for reference on each centrality plot.}
    \label{fig:polar}
\end{figure}

A synthetic network composed of six communities of 20 nodes, with additional connections added randomly between communities is shown in Fig. \ref{fig:polar}a. The pattern of connections repeats for each community, which is reflected in the similarity of LEC distributions. Fig. \ref{fig:polar}b demonstrates that LEC produces more consistent centrality results and more similar results to Known-Community Eigenvector Centrality (KCEC) for this synthetic network example. LEC emulating the eigenvector centrality distributions of the known-communities without \textit{a priori} knowledge of the community structure but with reference to the largest eigengap at $i=6$. Eigenvector (EC), PageRank (PR) and Principal Component Centrality (PCC) displaying greater variation and less alignment to KCEC. 

\begin{figure}[t!]
    \centering
    \includegraphics[width=1\textwidth]{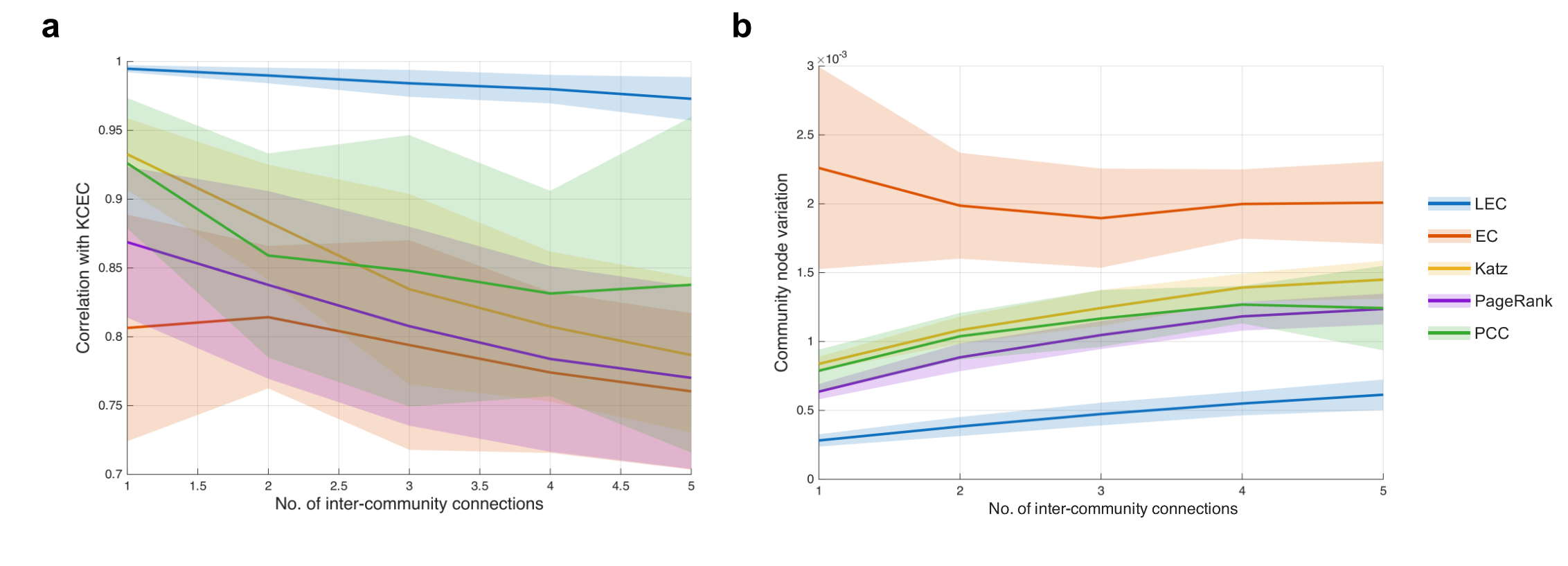}
    \caption{{\bf Centrality performance across variations of the six-community network.} Values reported as mean $\pm$ standard deviation from 10 networks at each level of inter-community connections. (a) Centrality correlation with KCEC. (b) Community node variation for each centrality across the six communities.}
    \label{fig:trends}
\end{figure}

The analysis is extended in Fig. \ref{fig:trends} to consider synthetic networks with randomised community configurations and additional variation in the number of inter-community connections, defined as the number of randomly allocated outgoing edges from each community. LEC is shown to be effective in identifying local centrality distributions with consistently higher correlations with KCEC in Fig. \ref{fig:trends}a.

There is no singular definition for centrality. Therefore, to account for variation in the design of centrality measures, Fig. \ref{fig:trends}b reports on the internal consistency of each centrality measure across the duplicated community topologies. LEC displays notably lower variation in centrality distribution within each duplicated community, with variation increasing as would be expected with the number of inter-community connections. 

There are a plethora of centrality measures, including those with seemingly similar intentions and remits as LEC. Additional comparisons include community-aware centrality measures that require \textit{a priori} knowledge of community structure \cite{rajeh2023comparative}. Despite having additional information, these methods were found to exhibit poor consistency across communities for the synthetic networks. Largely because they preferentially reward nodes that bridge communities in a manner that masks within-community influence. Community-based centrality \cite{zhao2015community} was also considered and produced similar correlations with KCEC and node variation results as Katz centrality. Finally, Neighborhood Centrality \cite{liu2016identify} and ClusterRank \cite{chen2013identifying} are prominent Neighbourhood and local centrality measures \cite{shvydun2025zoo}, but did not improve upon the performance of Katz, PageRank or PCC.

\begin{figure}[b!]
    \centering
    \includegraphics[width=.6\textwidth]{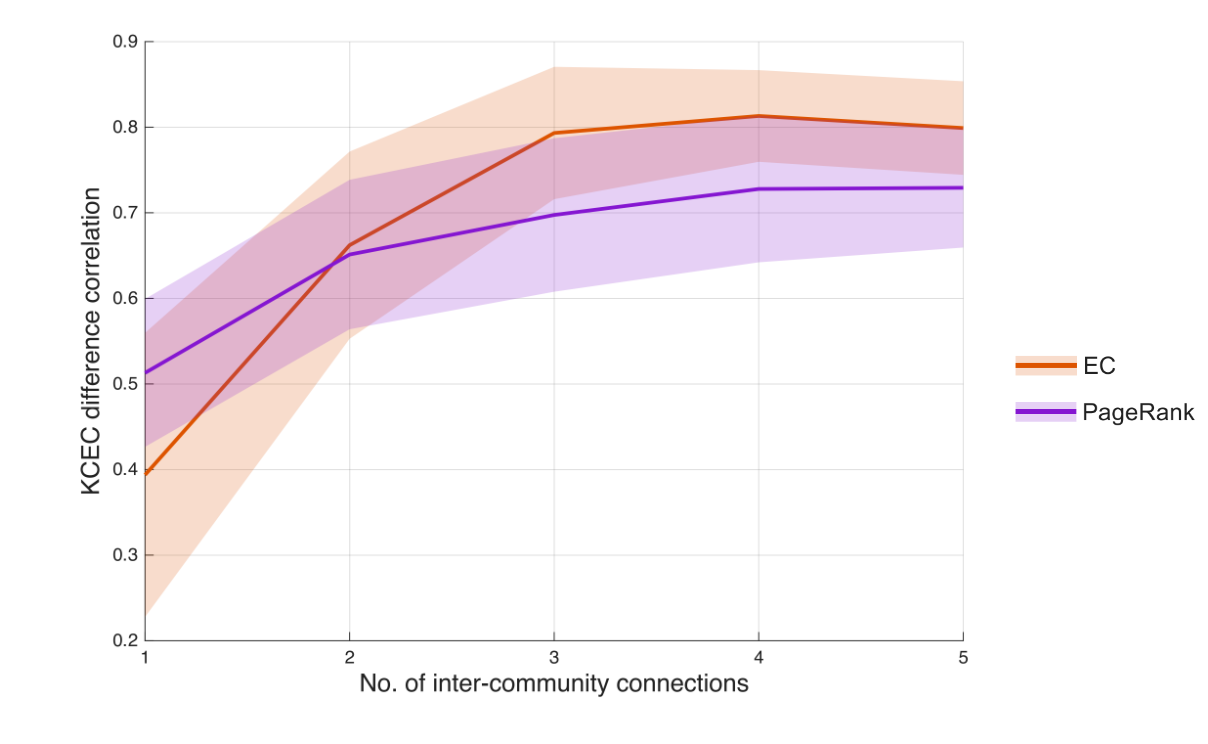}
    \caption{{\bf Spearman rank correlation of LEC$-$KCEC with EC$-$KCEC and PageRank$-$KCEC across variations of the six-community network.} Values reported as mean $\pm$ standard deviation from 10 networks at each level of inter-community connections.}
    \label{fig:corr_diff}
\end{figure}

To assess the detection of global influence, we compare LEC with established global centrality measures (EC and PageRank) for the same set of synthetic networks. Establishing the centrality contribution from global not local connectivity by calculating the difference in node centrality with respect to KCEC. Fig. \ref{fig:corr_diff} is consistent with LEC incorporating global connectivity information by showing high Spearman rank correlations between the LEC diffence values (LEC$-$KCEC) and the corresponding differences in global centrality measures (EC$-$KCEC \& PageRank$-$KCEC). Correlations are also seen to increase with more inter-community connections, which would increase the global connectivity for each network.

\begin{figure}[t!]
    \centering
    \includegraphics[width=1\textwidth]{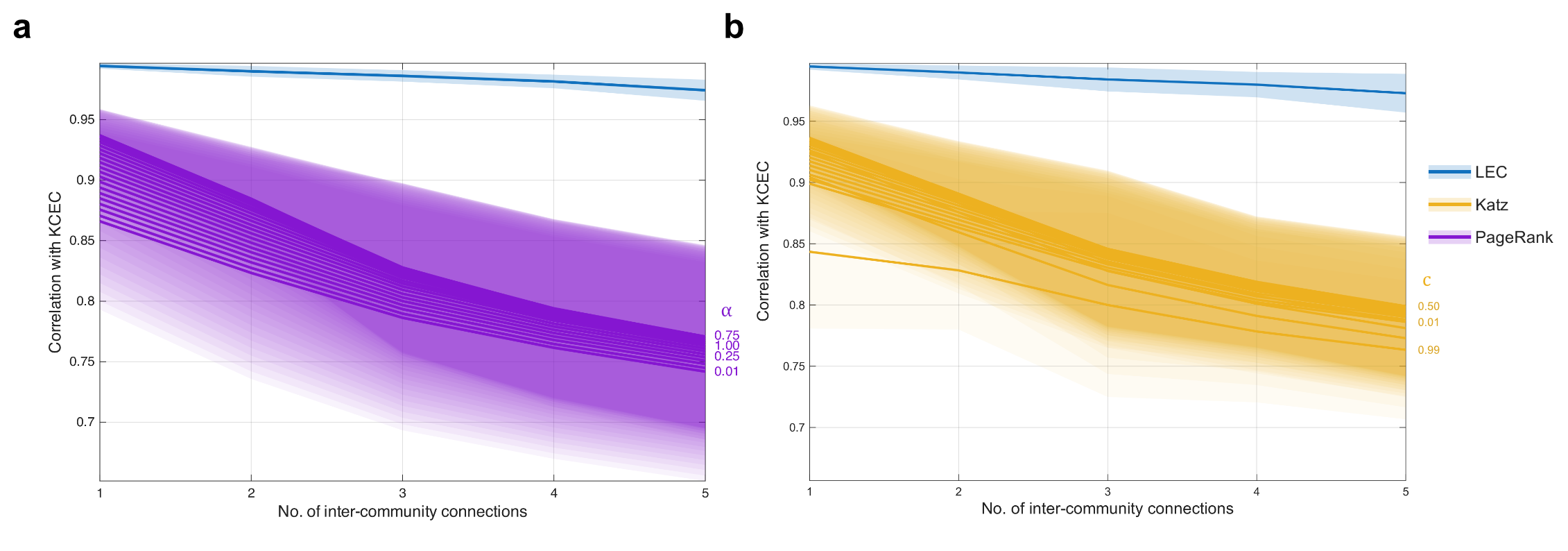}
    \caption{{\bf Centrality performance across variations of the six-community network.} Values reported as mean $\pm$ standard deviation from 10 networks at each level of inter-community connections. (a)     PageRank and LEC correlations with KCEC for probabilistic flow matrices. (b) Katz centrality and LEC correlations with KCEC for original adjacency matrices.}
    \label{fig:trends_PR}
\end{figure}

By transforming the synthetic network weights to represent probabilistic flow (see Section \ref{sec:hypo_testing}) we demonstrate in Fig. \ref{fig:trends_PR}a that the teleportation factor does not notably improve PageRank's correlation with KCEC. Note that when the adjacency matrix represents probabilistic flow and the teleportation factor is set as $\alpha=1$, PageRank performs the same assessment as eigenvector centrality. 
Similar to PageRank, Katz centrality has a damping factor $\alpha=c/\lambda_{\max}$ that affects the correlation with KCEC but Fig. \ref{fig:trends_PR}b shows that the performance never approaches that of LEC. Katz centrality is applied on the original adjacency matrix, with values approaching those of eigenvector centrality as $c\rightarrow 1$. These results suggest that adjusting the balance between short- and long-walk contributions is insufficient to recover the community informed centrality captured by LEC.

The community node variation of PageRank and Katz centralities is not visualised here as it does not make for a meaningful comparison. The variation is less than LEC for small values of $\alpha$ and $c$, however this performance is associated with a flattening of the centrality vector. Node scores approaching uniformity as the damping factor approaches 0. For PageRank, high teleportation rates reduce the influence of the network structure, while for Katz centrality, small damping factors heavily down-weight all but the shortest walks. Consequently, the reduced variation reflects a loss of structural information rather than improved community-aware centrality mapping

\subsection{Real-world networks}

School contact networks are real-world examples that contain known-communities corresponding to year and class structure with students spending more time with peers and classmates. Primary school \cite{Stehl2011} and high school \cite{MastrandreaR2015} contact networks provide real-world validation of the synthetic network results.

\subsection{Primary school}

\begin{figure}[t!]
    \centering
    \includegraphics[width=1\textwidth]{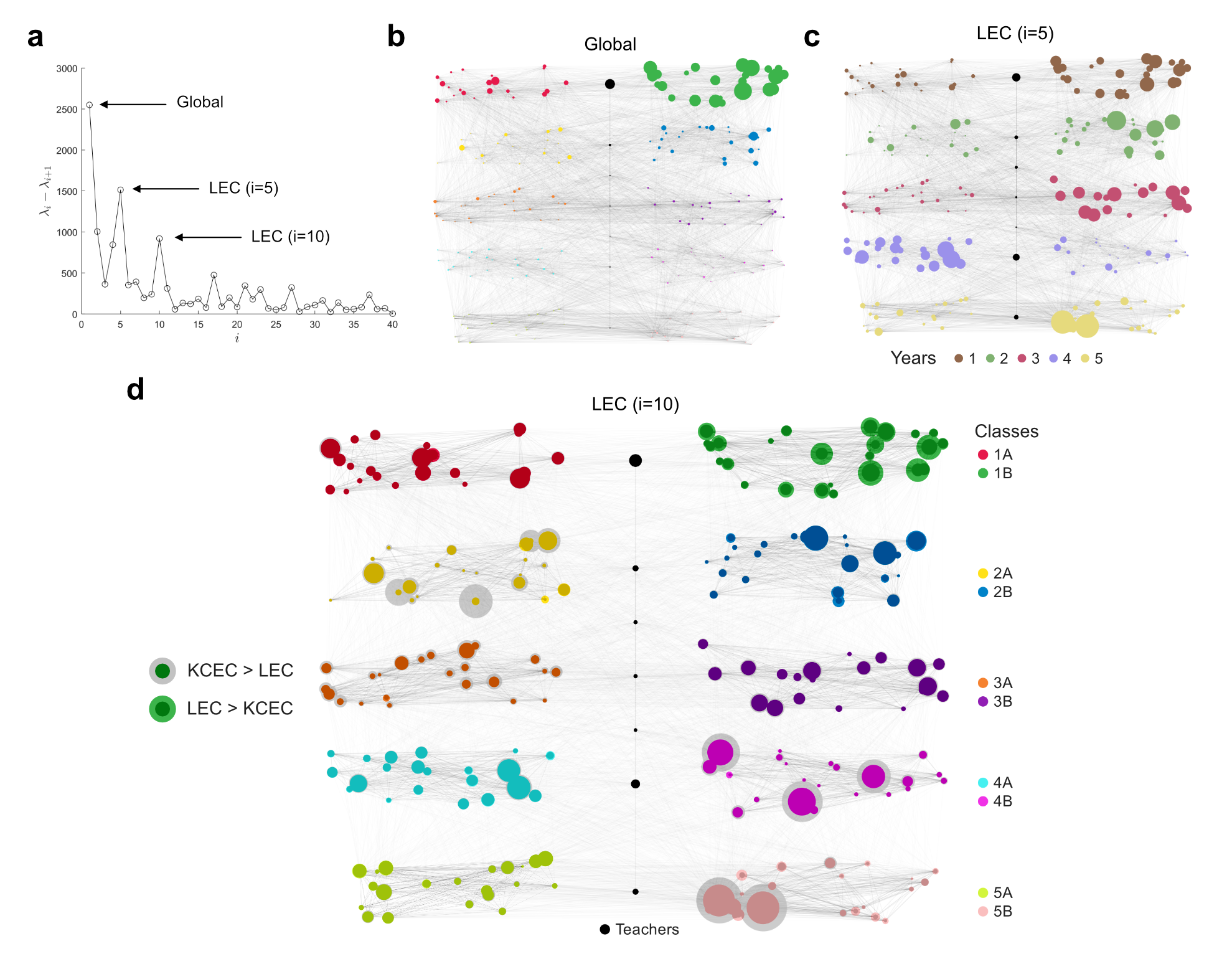}
    \caption{{\bf Centrality analysis of the primary school contact network.} (a) Eigengap plot showing prominent eigengaps at $i=1$,$5$, and $10$. (b) Eigenvector centrality distribution capturing global centrality. (c) LEC for $i=5$. (d) Comparison of LEC with KCEC for $i=10$. In panels (b–d), the marker sizes are proportional to the corresponding centrality values.}
    \label{fig:primary_school}
\end{figure}

For the primary school network, Figure \ref{fig:primary_school}a displays that the most prominent eigengap is associated with global connectivity ($i=1$). The high number of cross-class and year connections visible in Figure \ref{fig:primary_school}b,c,d delivering global connectivity. Class 1B is shown to be the most globally central and well-connected class in Figure \ref{fig:primary_school}b according to EC. 
There are also prominent eigengaps at $i=5$ and $i=10$ that align with the 5 year-groups and 10 classes. LEC ($i=5$) in Figure \ref{fig:primary_school}c displays a greater spread of centrality than EC with high value nodes in each of the five year groups. The expansion to LEC ($i=10$) in Figure \ref{fig:primary_school}d spreads the centrality across all 10 classes with high centrality nodes in all classes.

LEC ($i=10$) is compared with KCEC in Figure \ref{fig:primary_school}d, where the ten known -communities are defined as the ten primary school classes. 
The correlation between LEC ($i=10$) and KCEC for nine of the classes is high, with $c$ ranging from $0.90$ to $0.99$ for each class. The exception is Class 2A with $c=0.55$. However, Class 2A also has the lowest within-community contact durations (edge weights) with 31 hours in total, with a mean within-community contact duration of 50 hours across the other classes. Adjusting the network, by scaling only Class 2A's within-community connections weights by 1.5, results in a high correlation between LEC ($i=10$) and KCEC for Class 2A with $c=0.98$. 

The Primary school data \cite{Stehl2011} contains a second day of contact tracking. For this day, $i=10$ is again a prominent eigengap with seven of the classes reporting correlations $c$ from $0.90$ to $0.99$. The reduced correlations for three of the classes ($c=0.36, 0.45,0.68$) are again associated with the lowest within-community contact duration totals of 18, 32, 30 hours respectively, where the mean total across other classes is 63.5 hours. 
Adjusting the network through scaling the within-community weights by 2 (to account for the larger variation in within-community totals) increases $c>0.99$ for the three reduced correlation classes.

\subsection{Secondary school: community misalignment}
\begin{figure}[t!]
    \centering
    \includegraphics[width=.9\textwidth]{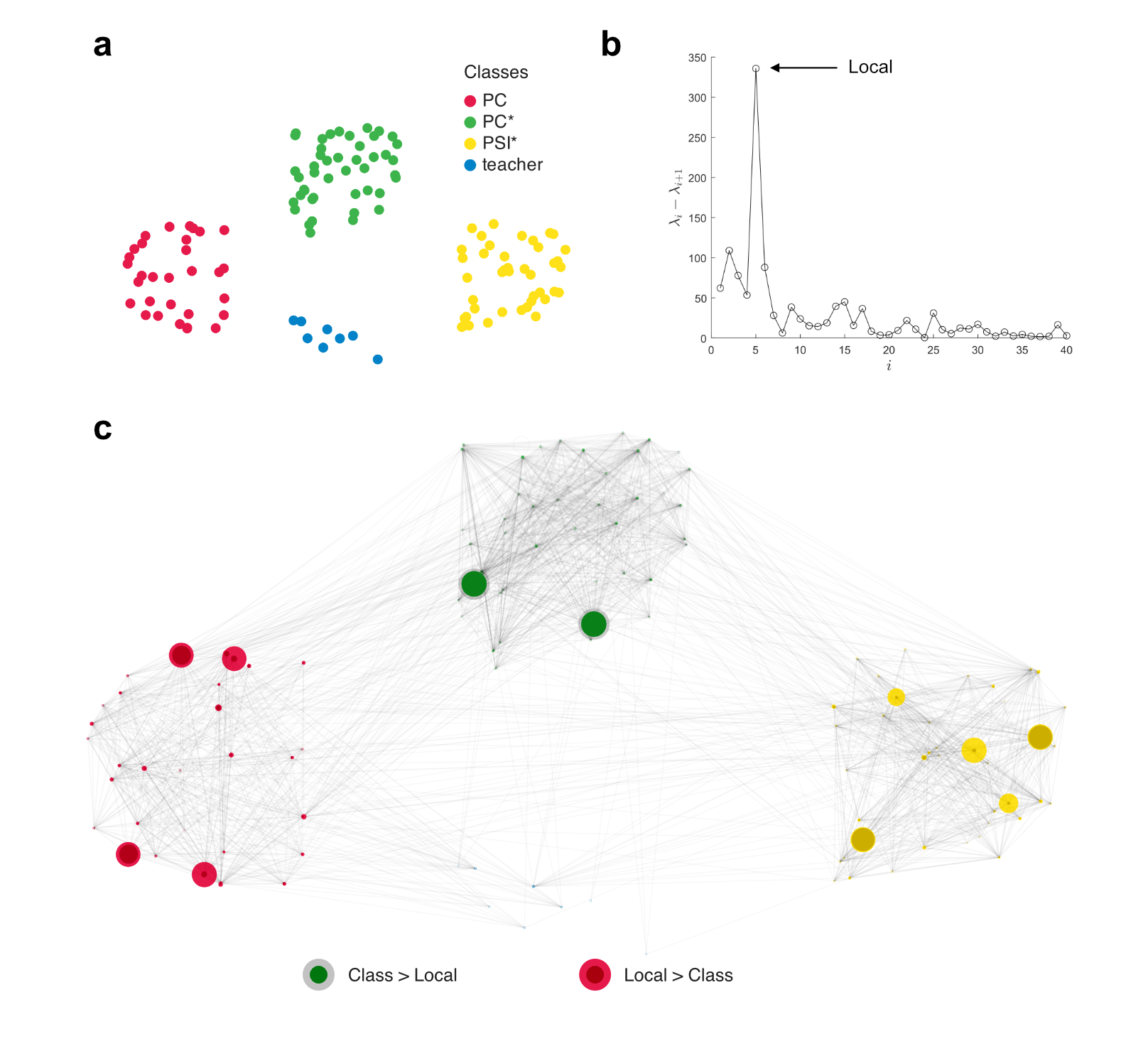}
    \caption{{\bf Centrality analysis of a high school contact network.} (a) Visualistion of class membership. (b) Eigengap plot showing prominent eigengap at $i=5$. (c) Comparison of LEC with KCEC for $i=5$. Circle sizes are proportional to the corresponding centrality values, illustrating the relative centrality of nodes according to these measures.}
    \label{fig:high_school}
\end{figure}

A high school contact network, \cite{MastrandreaR2015}, is analysed to reveal a mismatch between the number of defined classes (PC, PC*, and PSI*) and the largest eigengap, $i=5$, in Figure \ref{fig:high_school}.
 %
 This mismatch manifests as a large difference between LEC and KCEC for specific nodes in two of the classes in Figure \ref{fig:high_school}c. If LEC is constructed from only three of the first five eigenvectors ($v_1,v_2,v_4$) then $c=0.999$ for the correlation with KCEC. The second and fifth eigenvectors are found to correspond with sub-communities in two of the classes. The largest eigengap for both class' subnetworks is $i=2$, which indicates prominent division within both classes that eigenvector centrality does not reflect.

\subsubsection{Localisation mitigation}
\begin{figure}[t!]
    \centering
    \includegraphics[width=.8\textwidth]{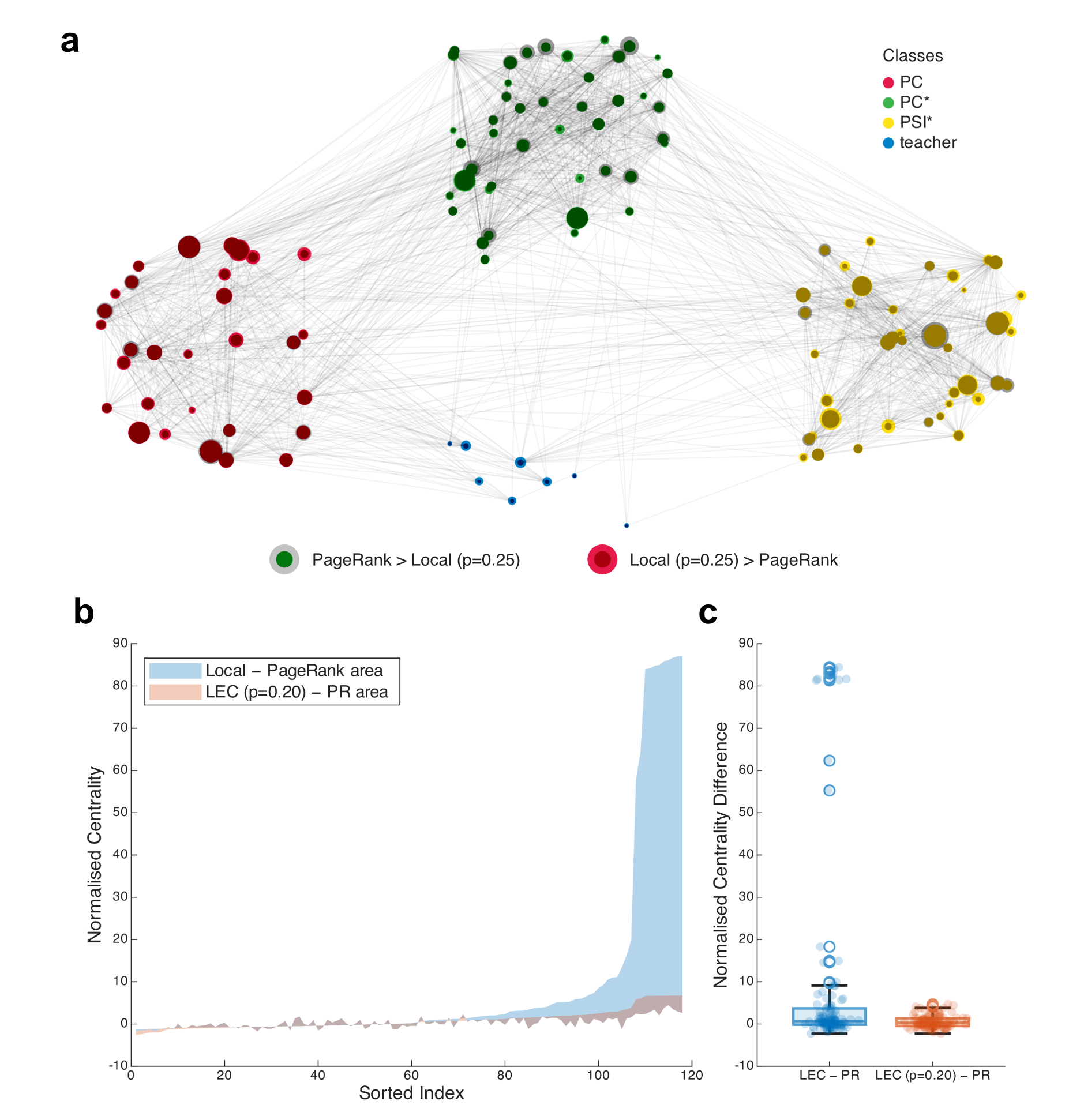}
    \caption{{\bf Localisation mitigation approach applied to High School contact network.} Comparison of (a) PageRank with LEC (i=5) where markers are proportional to centrality values. LEC, without and with p=0.2 non-linear rescaling, comparison with PageRank (PR) displayed as (b) difference area plot and (c) boxplot.}
    \label{fig:high_school_rescaling}
\end{figure}

Localisation limits the application of eigenvector-based centrality measures. For the high school network, localisation results in a skewed distribution for eigenvector-based centrality. PageRank provides a notably different assessment of centrality than LEC and KCEC, where it avoids localisation through the teleportation function and therefore distributes centrality more evenly throughout the network in Figure \ref{fig:high_school_rescaling}a. 
As previously demonstrated, 
PageRank does not provide an assessment of community-level centrality but it does assess global connectivity of nodes within all communities and therefore provides an effective reference for rescaling eigenvector-based centrality measures.
The non-linear rescaling of LEC, $x$, is defined as 
\begin{align}\label{3.1}
\mathcal{N}_p\left(x\right)=\frac{x^{\circ p}}{\sum{x^{\circ p}}}
\end{align}
where $x^{\circ p}$ is an element-wise Hadamard power. Figure \ref{fig:high_school_rescaling}a displays how rescaling of LEC can produce a similar distribution to PageRank. 
The minimum power ($p = 0.2$) was determined by minimising the Euclidean distance, $d$, with respect to PageRank, $z$. 
The initial correlation between LEC ($i=5$) and PageRank was 0.69. Following rescaling with $p=0.2$, this increases to $c=0.79$.

\section{Discussion} 
The results support local eigenvector centrality as a community-scale centrality measure that integrates global and local centrality, as summarised in Table \ref{table1}. The results on synthetic networks clearly distinguish LEC from the closely related spectral method Principal Component Centrality (PCC) \cite{Ilyas2010}, as well as from global centrality measures. PCC was the best performing alternative in terms of correlation with known-community eigenvector centrality (KCEC). However, PCC is unable to reproduce community-level centrality assessments due to a strong dependence on alignment with the dominant global spectral modes. Driven by the use of eigenvalues to weight contributing eigenvectors and the selection of contributing eigenvectors in reference to eigenvector centrality (EC).
\begin{table}[!t]
\caption{
{\bf Evidence supporting defined hypotheses.}}
\begin{tabular}{|m{4mm}|m{40mm}|m{75mm}|}
\hline
& \textbf{Definition} & \textbf{Evidence} \\
        \thickhline
        H1 & LEC is community-aware & LEC exhibits a strong correlation with community-level eigenvector centrality (KCEC) across both synthetic and real-world networks. Furthermore, LEC displays little variation in centrality among duplicated community topologies in the synthetic networks. \\ \hline
        H2 & LEC integrates global and local centrality & Deviations between LEC and KCEC are associated with increased global influence, which is consistent with LEC integrating information on global connectivity. \\ \hline
        H3 & Damping-based locality is not equivalent to community centrality & Varying the damping parameters of PageRank and Katz centrality alters how node ranking are affected by network topology, but neither method reproduces the community-scale centrality assessments obtained using LEC. \\ \hline
\end{tabular}
\label{table1}
\end{table}

Centrality is always considered within a bounded sphere of influence. The application of global centrality measures, such as EC and PageRank, implicitly assumes that node importance should consider connectivity to the whole network. In complex networks, it is often not clear whether node centrality is best judged at a global or local scale. The example of school contact networks can be studied across scales with notable global connectivity, as well as high connectivity within year groups and classes \cite{Stehl2011}. Without \textit{a priori} knowledge of the class structure, EC as a measure of global influence points towards the dominant presence of class 1B in Figure \ref{fig:primary_school}b with no central actors within years 4 and 5. Supported by an understanding of school structure, it would be expected that each class and year group has locally central nodes. LEC can through reference to prominent eigengaps capture centrality distributions at both a year and class scale. The efficacy of LEC in detecting local centrality demonstrated through comparison with KCEC.

An eigengap assessment is a key step in determining LEC. As noted by \cite{Shen2010,VonLuxburg2007}, the presence of the largest eigengap at an index greater than 1 indicates prominent community structure. The primary school example, displayed in Figure \ref{fig:primary_school}, demonstrates how a more nuanced picture of community structure emerges by investigating multiple prominent eigengaps that respond to known community structures. The largest eigengap is located at $i=1$, but prominent eigengaps are also found at $i=5$ and $i=10$ that reveal centrality distributions that respond to year groups and classes. Interrogation of prominent eigengaps is beyond the largest eigengap is particularly relevant in real-world networks where edge weightings implicitly capture relationships between nodes. For instance, the school contact networks weights edge linearly based on time in close proximity. A more representative weighting could employ nonlinear weighting as in \cite{Clark2021} or a more targeted measure such as time spent actively interacting. Such variations could affect the location of the largest eigengap.

The adjacency eigenvectors, selected according to prominent eigengaps, are found to reflect known structures in the graph for the school contact networks. Clustering using the adjacency matrix has been found previously to be most effective when employing a larger set of eigenvectors than would be suggested by the largest eigengap \cite{Luciska2018}. This suggests that not all communities will be prominently represented by a given LEC assessment. For example, less well-connected communities, such as the classes with lower total time in contact, may have their centrality reduced or influenced by connections external to their community. In this way, LEC is not a community detection or clustering algorithm, but an assessment of centrality that integrates local and global connectivity.

The localisation of eigenvectors diminishes their usefulness as centrality measures. PageRank \cite{Page1999} is used here as a reference measure for correcting localisation in LEC through non-linear rescaling. The damping factor that mitigates localisation for PageRank is also shown to not enable community-level centrality detection. This is consistent with the spectral interpretation of PageRank and Katz centrality as the dominant eigenvector of a perturbed matrix \cite{vigna2016spectral}. Although damping alters the extent to which rankings depend on network topology, it does not incorporate the mesoscopic spectral structure captured by LEC. 

\section{Conclusion}
Local eigenvector centrality (LEC) is a useful expansion on the established eigenvector centrality (EC); integrating locally prominent community structures with awareness of global connectivity. As a centrality measure, it provides a distinctive contribution through an adaptive assessment of node importance that can be tailored to prominent mesoscopic organisational scales within a network. Comparison with EC applied on known-communities in isolation demonstrates that eigenvectors beyond the dominant mode contain accurate insights into community-level centrality.

The utility of LEC in looking beyond global connectivity to identify central individuals is demonstrated for school contact networks that incorporate prominent year-group and class community structures. 
While the synthetic network experiments highlight the inability of alternative centrality heuristics to provide consistent community-level centrality assessments. 

By identifying a topologically relevant mesoscopic scale through prominent eigengaps, LEC provides an interpretable and effective integration of community-level and global centrality assessments. These results demonstrate that spectral centrality can be extended beyond dominant eigenvector analysis to accurately capture the underlying community organisation. 

\section{Data availability}
Code and example datasets used in this study are currently available in a GitHub repository: \url{https://github.com/RuaridhClark/Local_eigenvector_centrality}. The repository contains MATLAB and Colab notebooks, as well as the datasets analysed in the paper. The code and data will be archived on Zenodo following paper acceptance to obtain a permanent DOI.

\section*{Supporting information}

\paragraph*{S1 Appendix.}
\label{S1_Appendix}
\subsubsection{Complex Conjugates}
\begin{figure}[!t]
    \centering
    \includegraphics[width=.8\textwidth]{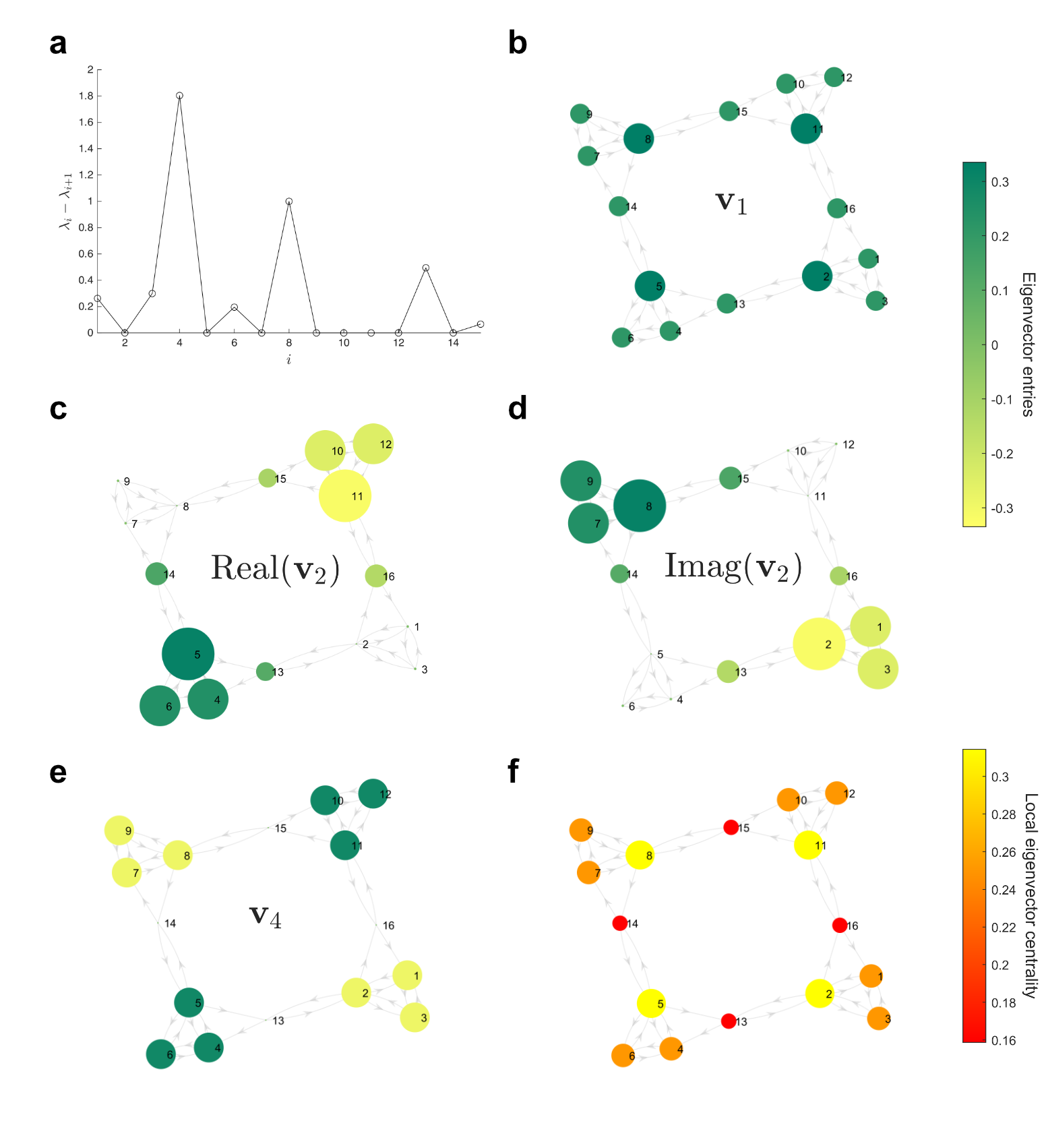}
    \caption{A four community graph example, where the complex conjugate pair occur at $v_2$ and $v_3$, with the real and imaginary values of $v_2$ displayed. (a) Eigengap analysis, (b) first eigenvector $v_1$, (c) absolute real component of $v_2$, (d) absolute imaginary component of $v_2$, (e) absolute $v_4$ values, and (f) local eigenvector centrality.}
    \label{fig:four-community}
\end{figure}

Complex conjugate pairs are rationalised in terms of centrality and community structure by considering the role of eigenvectors in decomposing a graph into a number of components/communities. For instance, the third eigenvector $v_3$ identifies decomposition into three separate node groupings when three valid groupings exist in a graph. However, when there are multiple valid configurations for allocating nodes into $\ell$ components, and/or no clear decomposition into $\ell$ components, then complex conjugates arise. Such a case is displayed in Figure \ref{fig:four-community}, where four communities are defined and visible in this toy example. The largest eigengap corresponds with $i=4$, as depicted in Figure \ref{fig:four-community}a, and the LEC result identifies equally prominent nodes in each community as shown in Figure \ref{fig:four-community}f. Note that the largest eigengap when taking the modulus of the eigenvalue differences results in $i=7$. 

The first and fourth eigenvectors of the four community example have only real eigenvalues and eigenvectors, displayed in Figure \ref{fig:four-community}b and e. Since there is no valid decomposition of the graph into three communities, the second and third eigenvectors are complex conjugate pairs, whereby the real components of both eigenvectors identify one configuration of dividing the graph into two components and the imaginary value components identify the other configuration, see Figure \ref{fig:four-community}c and d. 

The complex conjugate pairs are identified and replaced with two vectors based on the normalised real and imaginary component of one of the complex eigenvectors. Note that only the relative sign of the eigenvector entries is important, so either of the complex conjugate eigenvectors can be used to generate the two replacement vectors. 

\subsubsection{Directed Path}\label{sec:dir_path}

A directed path presents an edge case that is impacted by the alteration of the zero vector when assessing LEC. This directed acyclic graph takes the following form:

\begin{align}\label{2.10}
A = 
\begin{bmatrix}
0 & 1 & 0 & \cdots & 0 \\
0 & 0 & 1 & \cdots & 0 \\
\vdots & \vdots & \vdots & \ddots & \vdots \\
0 & 0 & 0 & \cdots & 1 \\
0 & 0 & 0 & \cdots & 0
\end{bmatrix}.
\end{align}

In this configuration, there is a single eigenvalue of $A$ that is equal to zero.  
Since an eigenvalue of zero indicates that the associated eigenvector corresponds to a direction of null transformation under the adjacency operator, this eigenvector offers no useful information for centrality analysis. Therefore, there is no centrality analysis possible in this example, and in the more general case, eigenvectors corresponding to zero eigenvalues should be excluded by setting their entries to zero.

\subsubsection{Directed Cycle}\label{sec:dir_cycle}

\begin{figure}[!hb]
    \centering
    \includegraphics[width=0.6\textwidth]{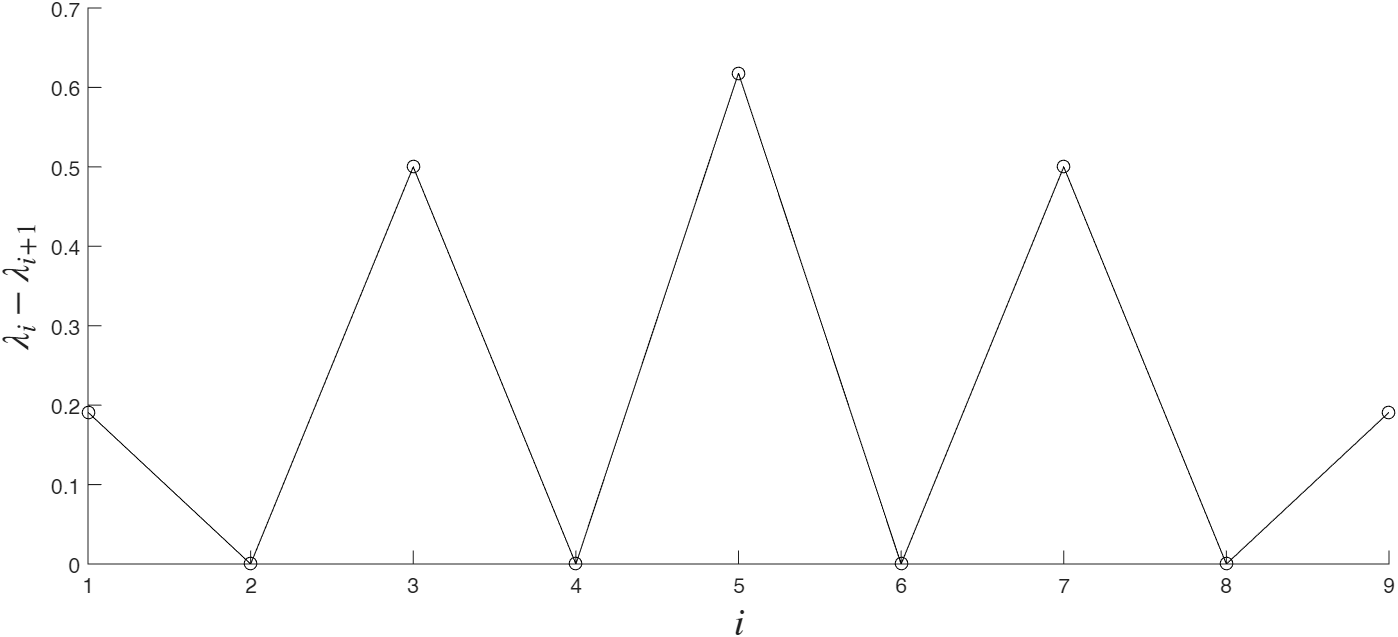}
    \caption{Eigengaps for a 10-node directed cycle}
    \label{fig:directed-cycle}
\end{figure}

A directed cycle provides another edge case for evaluating the proposed LEC. A directed cycle takes the following form:

\begin{align}\label{2.11}
A = 
\begin{bmatrix}
0 & 1 & 0 & \cdots & 0 \\
0 & 0 & 1 & \cdots & 0 \\
\vdots & \vdots & \vdots & \ddots & \vdots \\
0 & 0 & 0 & \cdots & 1 \\
1 & 0 & 0 & \cdots & 0
\end{bmatrix},
\end{align}

\noindent where the eigenvectors are complex conjugate pairs, except for the largest eigenvalue in a graph of odd order and for both the largest and smallest eigenvalues in a graph of even order. Despite the apparent lack of community structure, the cyclic nature of the graph results in the largest eigengap being located centrally in the eigengap order, as depicted in Figure \ref{fig:directed-cycle} for a $10$-node directed cycle. However, the combination of eigenvectors (eigenvectors $1$ to $5$ for a $10$-node directed cycle) produces an LEC of equal values for directed cycles of varying size, which is also equal to the eigenvector centrality (i.e., $v_1$).  
Both centrality measures correspond to the intuition that all nodes have equal influence in this cyclical graph.

%
\bibliography{Refs}

\end{document}